\documentclass[fleqn,10pt]{wlscirep}
\usepackage{ulem}
\usepackage{cite}
\usepackage[font=small,labelfont=bf,
   justification=justified,
   format=plain]{caption}
\usepackage{comment}   

\newcommand{\ra}[1]{\renewcommand{\arraystretch}{#1}}

\newcommand{\beginsupplement}{%
        \setcounter{table}{0}
        \renewcommand{\thetable}{\arabic{table}}%
        \setcounter{figure}{0}
        \renewcommand{\figurename}{Supplementary Figure}
        \renewcommand{\tablename}{Supplementary Table}
        \renewcommand{\refname}{Supplementary References}
}

\title{Insightful classification of crystal structures using deep learning}

\author[1,*]{Angelo Ziletti}
\author[2]{Devinder Kumar}
\author[1]{Matthias Scheffler}
\author[1]{Luca M. Ghiringhelli}
\affil[1]{Theory Department, Fritz-Haber-Institut der Max-Planck-Gesellschaft, Faradayweg 4-6, D-14195 Berlin, Germany}
\affil[2]{University of Waterloo, 200 University Avenue West, N2L 3G1 Waterloo-Ontario, Canada}

\affil[*]{ziletti@fhi-berlin.mpg.de}

\begin{abstract}
\textbf{Computational methods that automatically extract knowledge from data are critical for enabling data-driven materials science.
A reliable identification of lattice symmetry is a crucial first step for materials characterization and analytics. Current methods require a user-specified threshold, and are unable to detect average symmetries for defective structures. 
Here, we propose a machine-learning-based approach to automatically classify structures by crystal symmetry.
First, we represent crystals by calculating a diffraction image, then construct a deep-learning neural-network model for classification.
Our approach is able to correctly classify a dataset comprising more than $100\,000$ simulated crystal structures, including heavily defective ones.
The internal operations of the neural network are unraveled through attentive response maps, demonstrating that it uses the same landmarks a materials scientist would use, although never explicitly instructed to do so. 
Our study paves the way for crystal-structure recognition of - possibly noisy and incomplete - three-dimensional structural data in big-data materials science.}
\end{abstract}

\begin{document}

\flushbottom
\maketitle
\thispagestyle{empty}

\section*{Introduction}
Crystals play a crucial role in materials science. In particular, knowing chemical composition and crystal structure - the way atoms are arranged in space - is an essential ingredient for predicting properties of a material\cite{Olson2000,Curtarolo2003,Fischer2006}.
Indeed, it is well-known that the crystal structure has a direct impact on materials properties\cite{nye1985physical}. Just to give a concrete example: in iron, carbon solubility (important for steel formation) increases nearly forty times going from body-centered-cubic (bcc) $\alpha$-Fe (ferrite) to face-centered-cubic (fcc) $\gamma$-Fe (austenite)\cite{smith2004foundations}. From the computational point of view, identification of crystal symmetries allows for example to construct appropriate $k$-point grids for Brillouin zone sampling, generate paths between high-symmetry points in band structure calculations, or identify distortions for finite-displacement phonon calculations.

Given the importance of atomic arrangement in both theoretical and experimental materials science, an effective way of classifying crystals is to find the group of all transformations under which the system is invariant; in three-dimensions, these are described by the concept of space groups\cite{Hahn2006}. 
Currently, to determine the space group of a given structure, one first determines the allowed symmetry operations, and then compare them with all possible space groups to obtain the correct label; this is implemented in existing symmetry packages such as FINDSYM\cite{Stokes2005}, Platon\cite{Spek2009}, Spglib\cite{Grosse-Kunstleve1999,Englert2013,Spglib2017}, and most recently the self-consistent, threshold-adaptive AFLOW-SYM\cite{Hicks2018}. For idealized crystal structures, this procedure is exact. But in most practical applications atoms are displaced from their ideal symmetry positions due to (unavoidable) intrinsic defects or impurities or experimental noise. 
To address this, thresholds need to be set in order to define how loose one wants to be in classifying (namely, up to which deviations from the ideal structures are acceptable); different thresholds may lead to different classifications (see for instance Table \ref{table:accuracy-comparison}).
So far, this was not a big problem because individual researchers were manually finding appropriate tolerance parameters for their specific dataset.

However, our goal here is to introduce an automatic procedure to classify crystal structures starting from a set of atomic coordinates and lattice vectors; this is motivated by the advent of high-throughput materials science computations, thanks to which millions of calculated data are now available to the scientific community (see the Novel Materials Discovery (NOMAD) Laboratory\cite{NOMAD} and references therein).
Clearly, there is no universal threshold that performs optimally (or even sub-optimally) for such a large number of calculations, nor a clear procedure to check if the chosen threshold is sound.
Moreover, the aforementioned symmetry-based approach fails  - regardless of the tolerance thresholds - in the presence of defects such as, for example, vacancies, interstitials, antisites, or dislocations.
In fact, even removing a single atom from a structure causes the system to lose most of its symmetries, and thus one typically obtains the (low symmetry, e.g. $P1$) space group compatible with the few symmetry operations preserved in the defective structure. This label - although being technically correct - is practically always different from the label that one would consider appropriate (i.e. the most similar space group, in this case the one of the pristine structure).
Robustness to defects, however, is paramount in local and global crystal structure recognition.
Grain boundaries, dislocations, local inclusions, heterophase interfaces, and in general all crystallographic defects can have a large impact on macroscopic materials properties (e.g. corrosion resistance\cite{Ryan2002,Duarte2013}).
Furthermore, atom probe tomography - arguably the most important source of local structural information for bulk systems - provides three-dimensional atomic positions with an efficiency up to 80\%\cite{Gault2012} and near-atomic resolution; which, on the other hand, means that at least 20\% of atoms escaped detection, and the uncertainty on their positions is considerable. 

Here, we propose a procedure to efficiently represent and classify potentially noisy and incomplete three-dimensional materials science structural data according to their crystal symmetry (and not to classify x-ray diffraction images, or powder x-ray diffraction data\cite{Park2017}). These three-dimensional structural data could be for example atomic structures from computational materials science databases, or elemental mappings from atom-probe tomography experiments.
Our procedure does not require any tolerance threshold, and it is very robust to defects (even at defect concentrations as high as 40\%).  First, we introduce a way to represent crystal structures (by means of images, i.e. two-dimensional maps of the three-dimensional crystal structures, see below), then we present a classification model based on convolutional neural networks, and finally we unfold the internal behavior of the classification model through visualization.
An interactive online tutorial for reproducing the main results of this work is also provided\cite{ziletti2018}.

\begin{figure}[ht]
\centering
\includegraphics[width=9.0cm]{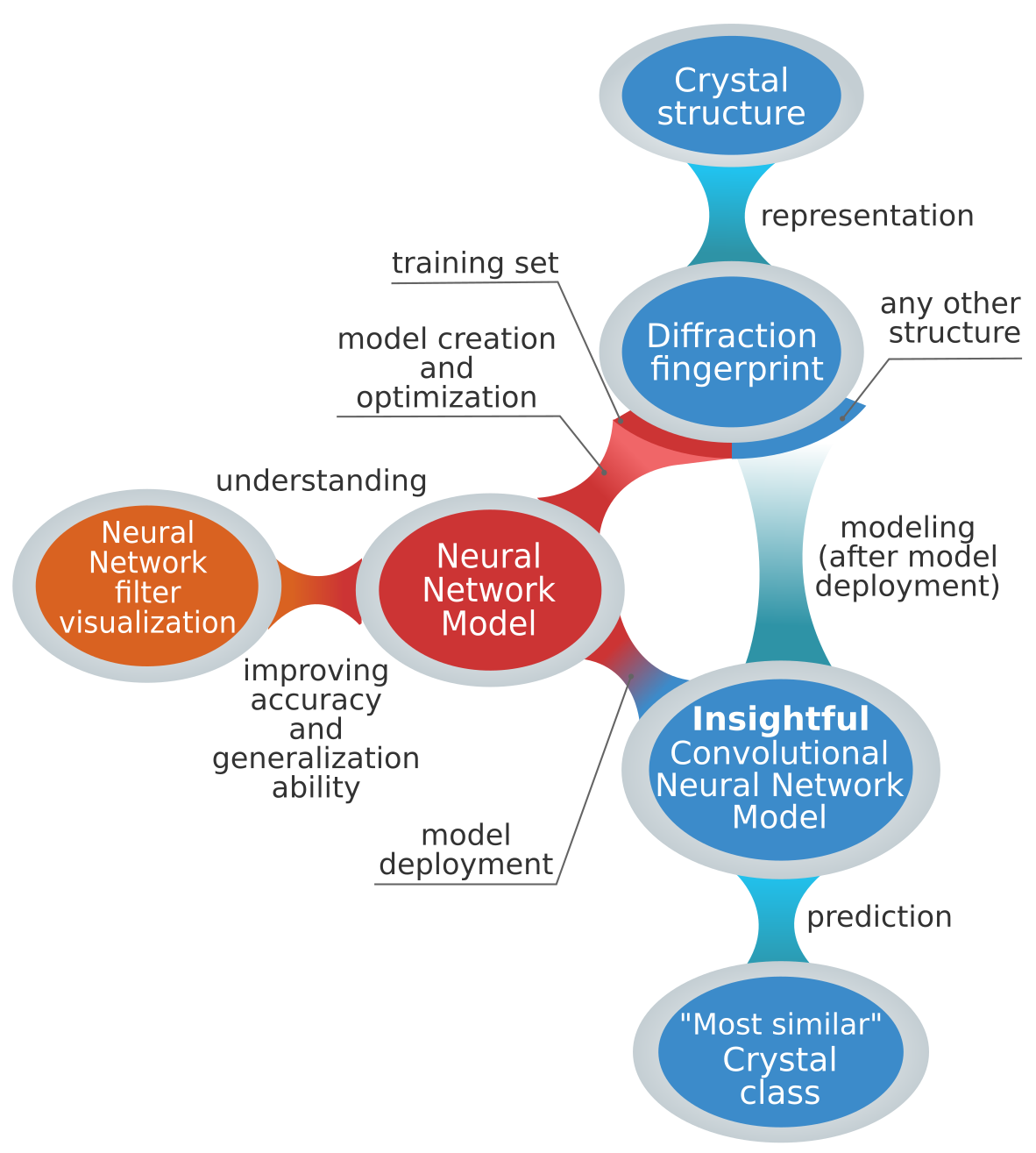}
\caption{The model workflow of automatic crystal-structure classification. First, every crystal structure is represented by the two-dimensional diffraction fingerprint. Then, a small subset of these structures are used as training set to generate a classification model. In particular, a convolutional neural network is used, and optimized minimizing the training set classification error. However, this is in general not enough to have a sound and generalizable model. Thus, we unfold the neural network internal operations by visualization, and ensure that the model arrives at its classification decision on physically motivated grounds. Finally, a classification model is deployed, and crystal structures can be directly and efficiently classified without any additional model optimization.}
\label{fig:method_pipeline}
\end{figure}

\section*{Results}

\subsection*{How to represent a material}

The first necessary step to perform any machine learning and/or automatized analysis on materials science data (see Fig. \ref{fig:method_pipeline}) is to represent the material under consideration in a way that is understandable for a computer. 
This representation - termed ``descriptor"\cite{Ghiringhelli2015} - should contain all the relevant information on the system needed for the desired learning task.
Numerous structural descriptors have been proposed to represent physical systems, most notable examples being atom-centered symmetry functions\cite{Behler2007}, Coulomb matrix\cite{Rupp2012}, smooth overlap of atomic positions\cite{Bartok2013}, deep tensor neural networks\cite{Schutt2017}, many-body tensor representation\cite{Huo2017}, and Voronoi tessellation\cite{Ward2017,Isayev2016}. However, these descriptors are either not applicable to extended systems\cite{Rupp2012, Schutt2017}, not size-invariant by construction\cite{Huo2017}, or base their representation of infinite crystals on local neighborhoods of atoms in the material\cite{Behler2007, Bartok2013, Zhu2016, Ward2017,Isayev2016}.
If on the one hand these local approaches are able to produce accurate force-fields \cite{Deringer2017,Morawietz2016}, on the other hand their strategy of essentially partitioning the crystal in patches (defined by a certain cut-off radius, generally 4-6~\text{\AA}\cite{Behler2007, Deringer2017}) makes it difficult to detect global structural properties, in particular where recognizing long-range order is crucial.

\begin{figure}[ht]
\centering
\includegraphics[width=18cm]{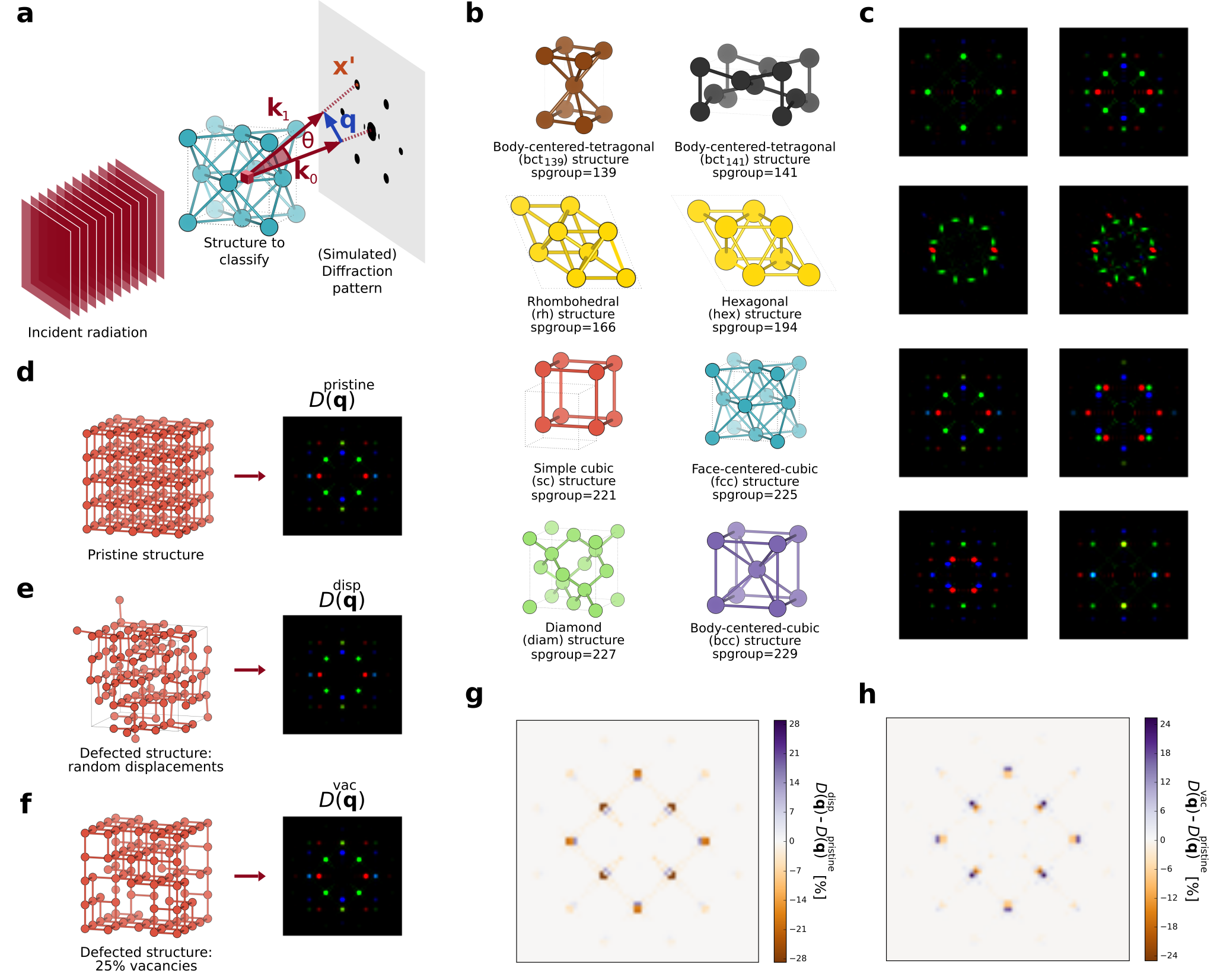}
\caption{The two-dimensional diffraction fingerprint. (a) Schematic representation of the two-dimensional diffraction fingerprint calculation. An incident plane wave is scattered by the material, and the diffraction pattern on a plane perpendicular to the incident radiation is computed. (b) Prototypes of the crystal classes considered in this work. (c) Examples of two-dimensional diffraction patterns for materials belonging to each of the eight classes. The ordering is the same as b. Rhombohedral and hexagonal structures have the same two-dimensional diffraction fingerprint.(d)-(e)-(f) A pristine simple-cubic structure (d), the same structure with 25\% of vacancies (e), and with atoms displaced randomly according to a Gaussian distribution with standard deviation of 0.08 ~\AA\ (f), together with their diffraction fingerprints. (g) (h) Difference between the diffraction fingerprints of the defective e-f and the pristine structure d.}
\label{fig:descriptor}
\end{figure}

In the case of crystal-structure recognition, however, it is essential that the descriptor captures system's symmetries in a compact way, while being size-invariant in order to reflect the infinite nature of crystals. Periodicity and prevailing symmetries are evident - and more compact - in reciprocal space, and therefore we introduce an approach based on this space.
For every system, we first simulate the scattering of an incident plane wave through the crystal, and then we compute the diffraction pattern in the detector plane orthogonal to that incident wave. This is schematically depicted in Fig. \ref{fig:descriptor}a.
The amplitude $\Psi$, which originates from the scattering of a plane wave with wave-vector $\mathbf{k}_0$ by $N_a$ atoms of species $a$ at positions $\lbrace \mathbf{x}_j^{(a)} \rbrace$ in the material can be written as:
\begin{align}
\label{eq:psi_q}
\Psi \left(\mathbf{q}\right) = r^{-1} \sum_a f_a^{\lambda}\left(\theta\right) \left[\sum_{j=1}^{N_a}r_0 \exp{\left( -i \mathbf{q} \cdot \mathbf{x}_j^{(a)} \right)} \right]
\end{align}
where $r_0$ is the Thomson scattering length, $\mathbf{q} = \mathbf{k}_1 - \mathbf{k}_0$ is the scattering wave-vector, $\mathbf{x}^\prime$ the corresponding position in the detector plane, and $r= \lvert \mathbf{x}^\prime \rvert$ (see Fig.\ref{fig:descriptor}a). Assuming elastic scattering, we have that $\lvert \mathbf{k}_0 \rvert = \lvert \mathbf{k}_1 \rvert = 2 \pi/ \lambda$, where $\lambda$ is the wavelength of the incident radiation. 
The quantity $f_a^{\lambda}\left(\theta\right)$ is the so-called x-ray form factor; it describes how an isolated atom of species \textit{a} scatters incident radiation with wavelength $\lambda$ and scattering angle $\theta$.
Since x-rays are scattered by the electronic cloud of an atom, its amplitude increases with the atomic number $Z$ of the element\cite{Henke1993}.
Following the successful application of scattering concepts in determining atomic structures (using for example x-rays\cite{Friedrich1913}, electrons\cite{THOMSON1927} or neutrons\cite{Wollan1948}), we propose the diffraction pattern intensity as the central quantity to describe crystal structures: 
\begin{align}
\label{eq:I_q}
I \left(\mathbf{q}\right) = A \cdot \Omega \left(\theta\right) \lvert \Psi \left(\mathbf{q}\right) \rvert^{2} 
\end{align}
where $\Omega \left(\theta\right)$ is the solid angle covered by our (theoretical) detector, and $A$ is a (inessential) constant determined by normalization with respect to the brightest peak (see section Methods). 
For each structure we first construct the standard conventional cell according to Ref.\cite{Setyawan2010}. Then, we rotate the structure $45^{\circ}$ clockwise and counterclockwise about a given crystal axis (e.g. $x$), calculate the diffraction pattern for each rotation, and superimpose the two patterns. Any other choice of rotation angle is in principle valid, provided that the diffraction patterns corresponding to different crystal classes do not accidentally become degenerate. This procedure is then repeated for all three crystal axes. The final result is represented as one RGB image for crystal structure, where each color channel shows the diffraction patterns obtained by rotating about a given axis (i.e. red (R) for $x$-axis, green (G) for $y$-axis, and blue (B) for $z$-axis). Each system is thus described as an image, and we term this descriptor two-dimensional diffraction fingerprint ($D_{\text{F}}$). 
We point out that this procedure does not require to already know the crystal symmetry, and $x$, $y$, and $z$ are arbitrary, e.g. determined ordering the lattice vectors by length\cite{Setyawan2010} (or whatever the chosen criterion). For additional computational details on the descriptor $D_{\text{F}}$, please refer to the section Methods.

Despite its rather complicated functional form (see Eqs. \ref{eq:psi_q} and \ref{eq:I_q}), the descriptor $D_{\text{F}}$ is one image for each system being represented (data point); the eight crystal classes considered in this work (see below) and examples of their calculated two-dimensional diffraction fingerprints are shown in Fig. \ref{fig:descriptor}b and Fig. \ref{fig:descriptor}c, respectively.
This descriptor compactly encodes detailed structural information (through Eq. \ref{eq:psi_q}) and - in accordance with scattering theory - has several desirable properties for crystal-structure classification, as we outline below. 

It is invariant with respect to system size: changing the number of periodic replicas of the system will leave the diffraction peak locations unaffected. This allows to treat extended and finite systems on equal footing, making our procedure able to recognize global and local order, respectively. 
We exploit this property, and instead of using periodically repeated crystals, we calculate $D_{\text{F}}$ using clusters of approximately 250 atoms. These clusters are constructed replicating the crystal unit cell (see Methods). By using finite samples, we explicitly demonstrate the local structure recognition ability of our procedure. 
The diffraction fingerprint is also invariant under atomic permutations: re-ordering the list of atoms in the system leads to the same $D_{\text{F}}$ due to the sum over all atoms in Eq. \ref{eq:psi_q}. 
Moreover, its dimension is independent of the number of atoms and the number of chemical species in the system being represented.
This is an important property because  machine learning models trained using this descriptor generalize to systems of different size by construction. This is not valid for most descriptors: for example, the Coulomb matrix dimension scales as the square of atoms in the largest molecule considered\cite{Rupp2012}, while in symmetry functions-based approaches\cite{Behler2007} the required number of functions (and thus model complexity) increases rapidly with the number of chemical species and system size.
Being based on the process of diffraction, the diffraction fingerprint mainly focuses on atomic positions and crystal symmetries; the information on the atomic species - encoded in the form factor $f_{a}^{\lambda}$ in Eq. \ref{eq:psi_q} - plays a less prominent role in the descriptor. As a result, materials with different atomic composition but similar crystal structure have similar representations. This is the ideal scenario for crystals classification: a descriptor which is similar for materials within the same class, and very different for materials belonging to different classes. Finally, the diffraction fingerprint is straightforward to compute, easily interpretable by a human (it is an image, see Fig. \ref{fig:descriptor}c), has a clear physical meaning (Eqs. \ref{eq:psi_q} and \ref{eq:I_q}), and is very robust to defects. This last fact can be traced back to a well-known property of the Fourier transform: the field at one point in reciprocal space (the image space in our case) depends on all points in real space. In particular, from Eq. \ref{eq:psi_q} we notice that the field $\Psi$ at point $\mathbf{q}$ is given by the sum of the scattering contributions from all the atoms in the system. If for example, some atoms are removed, this change will be smoothen out by the sum over all atoms and spread over - in principle - all points in reciprocal space. Practically, with increasing disorder new low-intensity peaks will gradually appear in the diffraction fingerprint  due to the now imperfect destructive interference between the atoms in the crystal. Examples of highly defected structures and their corresponding diffraction fingerprint are shown in Fig. \ref{fig:descriptor}e-\ref{fig:descriptor}f. It is evident that the diffraction fingerprint is indeed robust to defects. This property is crucial in enabling the classification model to obtain a perfect classification even in the presence of highly defective structures (see below).

A disadvantage of the two-dimensional diffraction fingerprint is that it is not unique across space groups. 
This is well-known in crystallography: the diffraction pattern does not always determine unambiguously the space group of a crystal\cite{Looijenga-Vos2006, de2007structure}. 
This is primarily because the symmetry of the diffraction pattern is not necessarily the same as the corresponding real-space crystal structure; for example, Friedel’s law states that - if anomalous dispersion is neglected - a diffraction pattern is centrosymmetric, irrespective of whether or not the crystal itself has a centre of symmetry. Thus, the diffraction fingerprint $D_{\text{F}}$ cannot represent non-centrosymmetric structures by construction. 
The non-uniqueness of the diffraction pattern $I \left(\mathbf{q}\right)$ across space groups also implies that crystal structures belonging to different space groups can have the same diffraction fingerprints.
Nevertheless, from Fig. \ref{fig:descriptor}c we notice that out of the eight crystal structure prototypes considered (covering the large majority of the most thermodynamically stable structures formed in nature by elemental solids\cite{ashcroft2011solid}), only the rhombehedral and hexagonal structures - whose real-space crystal structures are quite similar - have the same two-dimensional diffraction fingerprint.

\subsection*{The classification model}
Having introduced a way to represent periodic systems using scattering theory, we tackle the problem of their classification in crystal classes based on symmetries.
A first (and naive) approach to classify crystals - now represented by the diffraction descriptor $D_{\text{F}}$ - would be to write specific programs that detect diffraction peaks in the images, and classify accordingly.
Despite appearing simple at first glance, this requires numerous assumptions and heuristic criteria; one would need to define what is an actual diffraction peak and what is just noise, when two contiguous peaks are considered as one, how to quantify relative peak positions, to name but a few. 
In order to find such criteria and determine the associated parameters, one in principle needs to inspect all (thousands or even millions) pictures that are being classified. These rules would presumably be different across classes, require a separate - and not trivial - classification paradigm for each class, and consequently lead to a quagmire of ad-hoc parameters and task-specific software. In addition, the presence of defects leads to new peaks or alters the existing ones (see  Fig. \ref{fig:descriptor}g and \ref{fig:descriptor}h), complicating matters even further. Thus, this approach is certainly not easy to generalize to other crystal classes, and lacks a procedure to systematically improve its prediction capabilities.

\begin{figure}[ht]
\centering
\includegraphics[width=18cm]{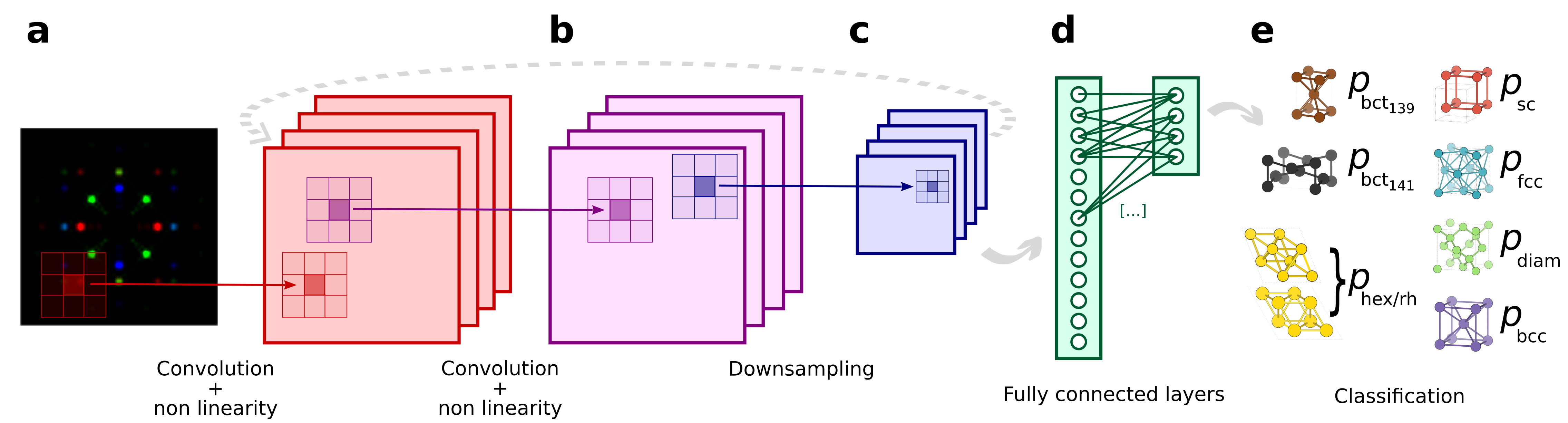}
\caption{Schematic representation of the convolutional neural network (ConvNet) used for crystals classification. (a) A learnable filter (also called kernel) is convolved across the image, and the scalar product between the filter and the input at every position is computed. This results in a two-dimensional activation map (in red) of that filter at each spatial position, which is then passed through a rectified linear unit (ReLu)\cite{icml2010_NairH10}. (b) The same procedure as point a is applied to this activation map (instead of the original image), producing another activation map (in purple). 
(c) A downsampling operation (in blue) is performed to coarse-grain the representation.
Six convolutional and two downsampling (max-pooling) layers are stacked sequentially (see Methods for additional details).
(d) The output of the convolutional/downsampling layers sequence is passed to fully-connected layers (regularized using dropout\cite{Srivastava2014}) to complete the classification procedure.
(e) The ConvNet outputs the probabilities that the input image, and therefore the corresponding material, belongs to a given class.
Minimizing the classification error, the above-mentioned filters are learned - through back-propagation\cite{Rumelhart1986} - and they will activate when a similar feature (e.g. edges or curves for initial layers, and more complex motifs for deeper layers) appears in the input.}
\label{fig:convnet_classification}
\end{figure}

However, it has been shown that all these challenges can be solved by deep-learning architectures\cite{Bengio2009,Schmidhuber2015,LeCun2015}. These are computational non-linear models sequentially composed to generate representations of data with increasing level of abstraction.
Hence, instead of writing a program by hand for each specific task, we collect a large amount of examples that specify the correct output (crystal class) for a given input (descriptor image $D_{\text{F}}$), and then minimize an objective function which quantifies the difference between the predicted and the correct classification labels.
Through this minimization, the weights (i.e. parameters) of the neural network are optimized to reduce such classification error\cite{Hinton2006,Hinton2006a}. In doing so, the network automatically learns representations (also called features) which capture discriminative elements, while discarding details not important for classification. This task - known as feature extraction - usually requires a considerable amount of heuristics and domain knowledge, but in deep learning architectures is performed with a fully automated and general-purpose procedure\cite{LeCun2015}. 
In particular, since our goal is to classify images, we use a specific type of deep learning network which has shown superior performance in image recognition: the convolutional neural network (ConvNet)\cite{LeCun1989,LeCun1998,Krizhevsky2012}.
A schematic representation of the ConvNet used in this work is shown in Fig. \ref{fig:convnet_classification}.
ConvNets are inspired by the multi-layered organization of the visual cortex\cite{Pamies2017}: filters are learned in a hierarchical fashion, composing low-level features (e.g. points, edges or curves) to generate more complex motifs. In our case, such motifs encode the relative position of the peaks in the diffraction fingerprint for the crystal classes considered, as we will show below.

\subsection*{The model performance}
For every calculation in the AFLOWLIB elemental solid database\cite{Curtarolo2012, Taylor2014}, we determine its space group using a symmetry-based approach\cite{Grosse-Kunstleve1999,Englert2013} as implemented by the Spglib code.
We then extract all systems belonging to centrosymmetric space groups which are represented with more than 50 configurations. This gives us systems with the following space group numbers: 139, 141, 166, 194, 221, 225, 227, and 229. For the case of elemental solids presented here, these space groups correspond to body-centered-tetragonal (bct, 139 and 141), rhombohedral (rh, 166), hexagonal (hex, 194), simple cubic (sc, 221), face-centered-cubic (fcc, 225), diamond (diam, 227), and body-centered-cubic (bcc, 229) structures.
This represents a rather complete dataset since it includes the crystal structures adopted by more than $80$\% of elemental solids under standard conditions.\cite{ashcroft2011solid}
It is also a challenging dataset because it contains $10\,517$  crystal structures comprising 83 different chemical species, cells of various size, and structures that are not necessarily in the most stable atomic arrangement for a given composition, or even at a local energy minimum. 
This last point in particular could potentially be a problem for the symmetry-based approach: when crystals are not in a perfect arrangement, it can fail in returning the correct labels. In fact, if atoms are slightly displaced from their expected symmetry positions, the classification could return a different space group because symmetries might be broken by this numerical noise. To avoid this, we include in the pristine dataset only systems which are successfully recognized by the symmetry-based approach to belong to one of the eight classes above, thus ensuring that the labels are correct.
We refer to the above as pristine dataset; the dataset labels are the aforementioned space groups, except for rh and hex structures, which we merge in one class (hex/rh) since they have the same diffraction fingerprint (see Fig. \ref{fig:descriptor}c).

\begin{table*}\centering
\ra{1.5}
\begin{tabular}{@{}c rrrrrr c rrrr@{}}\hline \hline
& \multicolumn{6}{c}{Random Displacements ($\sigma$)} & \phantom{i}& \multicolumn{4}{c}{Vacancies ($\eta$)} \\ 
\cline{2-7} \cline{9-12}
&  $0.001$\AA & $0.002$\AA & $0.005$\AA 
&  $0.01$\AA & $0.02$\AA & $0.06$\AA 
& \phantom{i} 
& $1$ \% & $2$ \% 
& $15$ \% 
& $25$ \% 
\\ \hline
Spglib (tight) & 0.00 & 0.00 & 0.00 & 0.00 & 0.00 & 0.00 
&& 0.02  & 0.00 & 0.00 & 0.00
\\
Spglib (medium) & 73.70 & 0.00 & 0.00 & 0.00 & 0.00 & 0.00 
&& 0.02  & 0.00 & 0.00 & 0.00
\\
Spglib (loose) & 99.99 & 99.99 & 99.99 & 75.22 & 0.00& 0.00 
&& 0.01  & 0.00 & 0.00 & 0.00
\\
\textbf{This work} & 100.00 & 100.00 & 100.00 & 100.00 & 100.00 & 100.00 
&& 100.00  & 100.00 & 100.00 & 100.00
\\
\hline \hline
\end{tabular}
\caption{Accuracy in identifying the correct (most similar) crystal class in the presence of defects. The defective structures are calculated randomly displacing atoms according to Gaussian distribution with standard deviation $\sigma$ (left), or removing $\eta$\% of the atoms (right). For details regarding the Spglib thresholds chosen see Supplementary Note 1. The accuracy values shown in the table are in percentage.}
\label{table:accuracy-comparison}
\end{table*}

We apply the workflow introduced here (and schematically shown in Fig. \ref{fig:method_pipeline}) to this dataset. For each structure, we first compute the two-dimensional diffraction fingerprint $D_{\text{F}}$; then, we train the ConvNet on (a random) 90\% of the dataset, and use the remaining 10\% as test set.
We obtain an accuracy of 100\% on both training and test set, showing that the model is able to perfectly learn the samples and at the same time capable of correctly classifying systems which were never encountered before.
The ConvNet model optimization (i.e. training) takes 80 minutes on a quad-core Intel(R) Core(TM) i7-3540M CPU, while one class label is predicted - for a given $D_{\text{F}}$ - in approximately 70~ms on the same machine (including reading time). 
The power of machine learning models lies in their ability to produce accurate results for samples that were not included at training. In particular, the more dissimilar test samples are from the training samples, the more stringent is the assessment of the model generalization performance. 
To evaluate this, starting from the pristine dataset, we generate heavily defective structures introducing random displacements (sampled from Gaussian distributions with standard deviation $\sigma$), randomly substituting atomic species (thus forming binaries and ternaries alloys), and creating vacancies.
This results in a dataset of defective systems, for some of which even the trained eyes of a materials scientist might have trouble identifying the underlying crystal symmetries from their structures in real space (compare for example, the crystal structures in Fig.\ref{fig:descriptor}d with \ref{fig:descriptor}e and \ref{fig:descriptor}f).

As mentioned in the Introduction and explicitly shown below, symmetry-based approaches for space group determination fail in giving the correct (most similar) crystal class in the presence of defects. 
Thus, strictly speaking, we do not have a true label to compare with. However, since in this particular case the defective dataset is generated starting from the pristine, we do know the original crystal class for each sample. Hence, to estimate the model generalization capability, we label the defective structures  with the class label of the corresponding pristine (parental) system. This is a sensible strategy given that displacing, substituting or removing atoms at random will unlikely change the materials' crystal class. 
Using the ConvNet trained on the pristine dataset (and labels from the pristine structures), we then predict the labels for structures belonging to the defective dataset. 
A summary of our findings is presented in Table \ref{table:accuracy-comparison}, which comprises results for $10\,517 \times (6+4) = 105\,170 $ defective systems; additional data are provided in Supplementary Note 1 and 2.

When random displacements are introduced, Spglib accuracy varies considerably according to the threshold used; moreover, at $\sigma \geq 0.02$ \AA\ Spglib is never able to identify the most similar crystal class, regardless of threshold used.
Conversely, the method proposed in this work always identifies the correct class up to $\sigma$ as high as $0.06$ \AA.
Similar are the results for vacancies: Spglib accuracy is $\sim 0$\% already at vacancies concentrations of 1\%, while our procedure attains an accuracy of 100\% up to 40\% vacancies, and greater than $97$\% for vacancy concentrations as high as 60\% (Table \ref{table:accuracy-comparison} and Supplementary Table 2).
Since no defective structure was included at training, this represents a compelling evidence of both the model robustness to defects and its generalization ability.

\begin{figure}[ht]
\centering
\includegraphics[width=18cm]{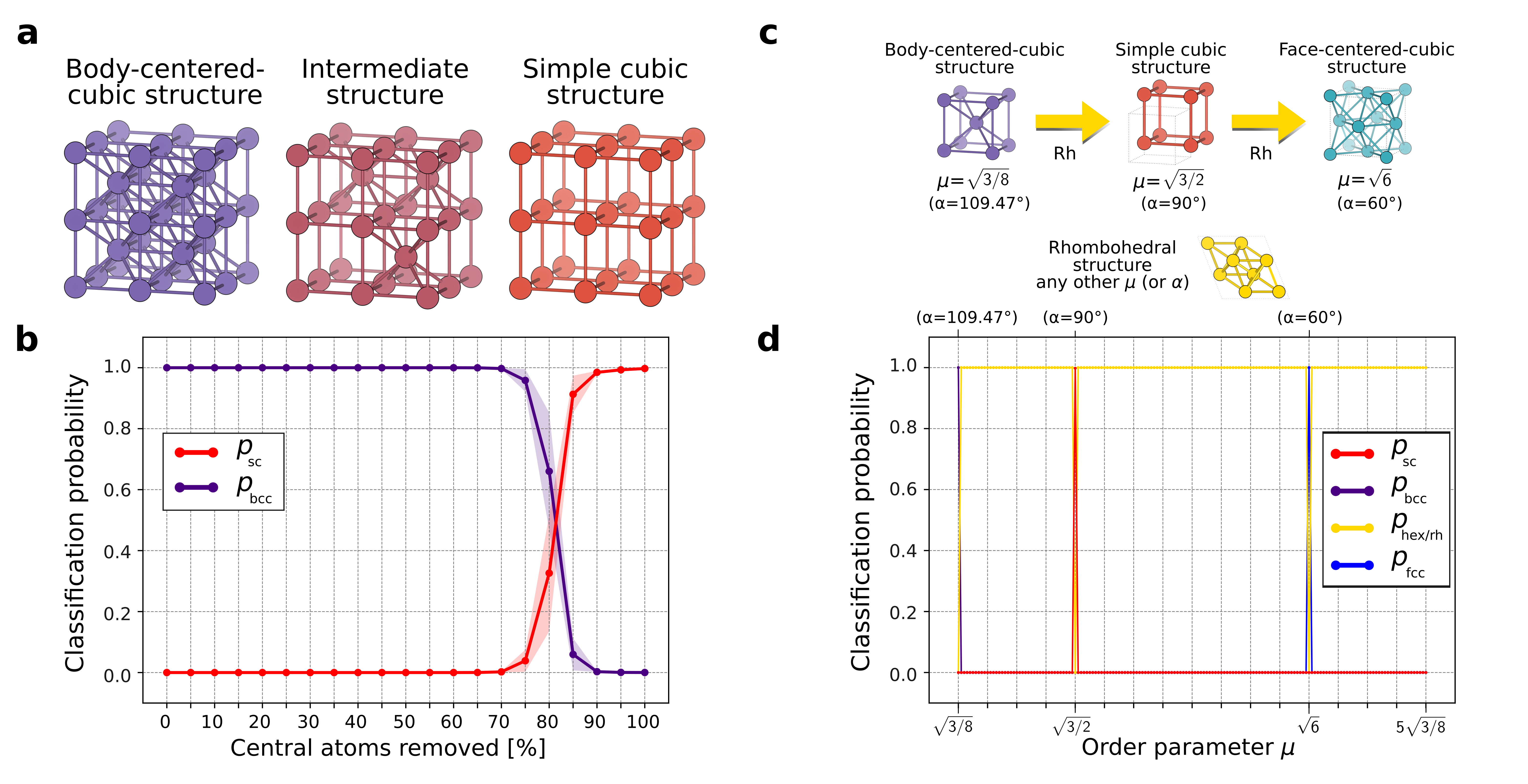}
  \caption{Neural network predictions on structural transitions. (a)(b) Body-centered-cubic (bcc) to simple cubic (sc) structural transition. (a) Examples of a bcc, an intermediate bcc/sc, and a sc structure. (b) Distributions of classification probability for the bcc (purple) and sc (red) classes as a function of the percentage of central atoms being removed (see text for more details). The shaded area corresponds to a range of one standard deviation above and below these distributions. 
(c)(d) Structural transition: transition path including rhombohedral, body-centered-cubic, simple-cubic and face-centered-cubic structures. 
The prototypes are generated using the AFLOW Library of Crystallographic Prototypes\cite{Mehl2016}.}
\label{fig:defects}
\end{figure}

If random changes will unlikely modify a crystal class, it is however possible to apply targeted transformations in order to change a given crystal from one class to another.
In particular, starting from a bcc one can obtain a sc crystal removing all atoms at the center of the bcc unit cell (Fig.\ref{fig:descriptor}b, and \ref{fig:defects}a). We remove different percentages of central atoms (from 0\% to 100\%, at 10\% steps) from a subset of bcc structures in the pristine dataset; this gives us a collection of structures which are intermediate between bcc and sc by construction (see Fig.\ref{fig:defects}a center for a concrete example).

Let us now recall that the output of our approach is not only the crystal class, but also the probability that a system belongs to a given class; this quantifies how certain the neural network is regarding its classification.
The probability of the aforementioned structures being fcc (purple) or sc (red) according to our model are plotted in Fig.\ref{fig:defects}b as function of the percentage of central atoms removed (the shaded area indicates the standard deviation of such distributions). 
This percentage can be seen as a order parameter of the bcc-to-sc structural phase transition.
If no atoms are removed, the structures are pure bcc, and the model indeed classifies them as bcc with probability 1, and zero standard deviation.
At first, removing (central) atoms does not modify this behavior: the structures are seen by the model as defective bcc structures .
However, at 75\% of central atoms removed, the neural network judges that such structures are not defective bcc anymore, but are actually intermediate between bcc and sc. This is reflected in an increase of the classification probability of sc, a corresponding decrease in bcc probability, and a large increment in the standard deviation of these two distributions. 
When all central atoms are removed, we are left with pure sc structures, and the model classifies again with probability 1, and vanishing standard deviation: the neural network is confident that these structures belong to the sc class. 

We conclude our model exploration applying the classification procedure on a structural transition path encompassing rhombohedral, body-centered-cubic, simple-cubic and face-centered-cubic structures. From the AFLOW Library of Crystallographic Prototypes\cite{Mehl2016}, we generate rhombohedral structures belonging to space group 166 (prototype $\beta$-Po A\_hR1\_166\_a) with different values of $\mu \equiv c/a$ or $\alpha$, where $a$ and $c$ are two of the lattice vectors of the conventional cell\cite{Setyawan2010}, and $\alpha$ is the angle formed by the primitive lattice vectors\cite{Mehl2016}.  
Particular values of $\mu$ (or $\alpha$) lead this rhombohedral prototype to reduce to bcc ($\mu_{\rm bcc}=\sqrt[]{3/8}$ or $\alpha=109.47^{\circ}$), sc ($\mu_{\rm sc}=\sqrt[]{3/2}$ or $\alpha=90^{\circ}$), or fcc ($\mu_{\rm fcc}=\sqrt[]{6}$ or $\alpha=60^{\circ}$) structures\cite{Mehl2016}. To test our model on this structural-transition path, we generate crystal structures with $\sqrt[]{3/8} \leq \mu \leq 5\sqrt[]{3/8}$, and use the neural network trained above to classify these structures. The results are shown in Fig. \ref{fig:defects}d. Our approach is able to identify when the prototype reduces to the high-symmetry structures mentioned above (at $\mu_{\rm bcc}$, $\mu_{\rm sc}$, and $\mu_{\rm fcc}$), and also correctly classify the structure as being rhombohedral for all other values of $\mu$. This is indeed the correct behavior: outside the high symmetry bcc/sc/fcc the structure goes back to hex/rh precisely because that is the lower symmetry family ($\mu$ not equal to $\mu_{\rm bcc}$, $\mu_{\rm sc}$, or $\mu_{\rm fcc}$).

\subsection*{Opening the black-box using attentive response maps}

Our procedure based on diffraction fingerprints and ConvNet correctly classifies both pristine and defective dataset but are we obtaining the right result for the right reason? And how does the ConvNet arrive at its final classification decision? 

To answer these questions, we need to unravel the neural network internal operations; a challenging problem which has recently attracted considerable attention in the deep learning community\cite{Zeiler2010,Zeiler2014,Bach2015,Kumar2016,Montavon2017,Kumar2017}.
The difficulty of this task lies in both the tendency of deep learning models to represent the information in a highly distributed manner, and the presence of non-linearities in the network's layers.
This in turn leads to a lack of interpretability which hindered the widespread use of neural networks in natural sciences: linear algorithms are often preferred over more sophisticated (but less interpretable) models with superior performance.

\begin{figure}[ht]
\centering
\includegraphics[width=18cm]{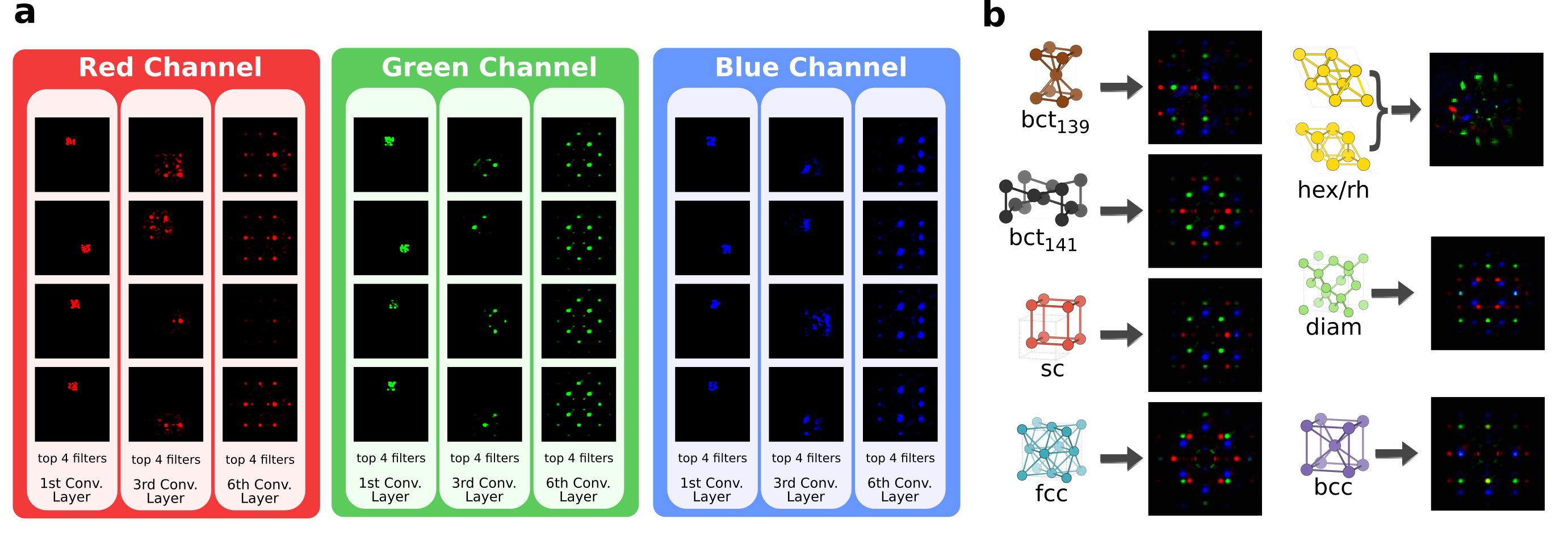}
\caption{Visualizing the convolutional neural network (ConvNet) attentive response maps. (a) Attentive response maps from the top four most activated filters of the first, third and last convolutional layers for the simple-cubic class. The brighter the pixel, the most important is that location for classification. Comparing across layers, we notice that the ConvNet filters are composed in a hierarchical fashion, increasing their complexity from one layer to another. At the third convolutional layer, the ConvNet discovers that the diffraction peaks, and their relative arrangement, are the most effective way to predict crystal classes. (b) Sum of the last convolutional layer filters for all seven crystal classes: the ConvNet learned crystal templates automatically from the data.}
\label{fig:convnet_interpretation}
\end{figure}

To shed light on the ConvNet classification process, we resort to visualization: using the fractionally strided convolutional technique introduced in Ref.\cite{Kumar2016} we back-projects attentive response maps (i.e. filters) in image space\cite{Zeiler2010,Zeiler2014,Kumar2017}.
Such attentive response maps - shown in Fig. \ref{fig:convnet_interpretation} - identify the parts of the image which are the most important in the classification decision\cite{Kumar2016}. 
The top four most activated (i.e. most important) filters from the first, third and last convolutional layers for each of the three color channels are shown in Fig. \ref{fig:convnet_interpretation}a for the sc class. 
The complexity of the learned filters grows layer by layer, as demonstrated by the increasing number of diffraction peaks spanned by each motif. The sum of the last convolutional layer filters for each class is shown in Fig. \ref{fig:convnet_interpretation}b; they are class templates automatically learned from the data by the ConvNet.
Comparing Fig.\ref{fig:descriptor}c and \ref{fig:convnet_interpretation}b, we see that our deep learning model is able to autonomously learn, and subsequently use, the same features that a domain expert would use. This not only confirms the soundness of the classification procedure, but also explains its robustness in terms of generalization.

\section*{Discussion}
We have introduced a way of representing crystal structures by means of (easily interpretable) images. Being based on reciprocal space, this descriptor - termed two-dimensional diffraction fingerprint - compactly encodes crystal symmetries, and possesses numerous attractive properties for crystal classification. In addition, it is complementary with existing real-space based representations\cite{Bartok2013}, making possible to envision a combined use of these two descriptors.
Starting from these diffraction fingerprints, we use a convolutional neural network to predict crystal classes. As a result, we obtain an automatic procedure for crystals classification which does not require any user-specified threshold, and achieves perfect classification even in the presence of highly defective structures.
On this regard, we argue that - since materials science data are generated in a relatively controlled environment - defective datasets represent probably the most suitable test to probe the generalization ability of any data-analytics model.
Given the solid physical grounds of the diffraction fingerprint representation, our deep learning model is modest in size, which translates in short training and prediction times.
Finally, using recently developed visualization techniques, we uncover the learning process of the neural network. Thanks to its multi-layered architecture, we demonstrate that the network is able to learn, and then use in its classification decision the same landmarks a human expert would use.
Further work is needed to make the approach proposed here unique across space groups and to widen its domain of applicability to non-centrosymmetric crystals, which can exhibit technologically relevant ferroelectric, piezoelectric or nonlinear optical effects.
In accordance with the principle of reproducible research\cite{Munafo2017,Baker2016}, we also provide an online tutorial\cite{ziletti2018} where users can interactively reproduce the main results of this work (but also produce their own) within the framework of the NOMAD Analytics-Toolkit.
As an outlook, our method could also be applied to the problem of local microstructure determination in atomic probe tomography experiments, with the ultimate goal of discovering structural-property relationships in real materials. 

\section*{Methods}

\subsection*{Two-dimensional diffraction fingerprint}
First, for each structure in the dataset (specified by a set of atomic coordinates and lattice vectors), we concatenate three random rotations around the three crystal axes to randomize the initial crystal orientation. Then, we construct the standard conventional cell according to Ref.\cite{Setyawan2010} using a customized implementation based on the Python Materials Genomics (pymatgen) package\cite{Ong2013}; in particular, we use the convention for triclinic cells - irrespective of the actual lattice type - and no symmetry refinement of the atomic position. This procedure is therefore completely independent from traditional symmetry approaches and robust against randomization of the initial crystal orientation. Finally, we replicate this standard cell in all three directions such that the resulting cluster contains a number of atoms which is as close as possible to a given target number (namely, 250). The size-invariance of the diffraction peak locations guarantees that the results are independent from this choice, only the peak widths will slightly change, in accordance with the indetermination principle\cite{sakurai2011modern} (this was expressly checked for systems ranging from 32 to 1024 atoms). Defective structures are then generated from these supercells by removing or randomly displacing atoms. We have also tested that a random rotation followed by the conventional cell determination applied to already generated defective structures leads to the same result, since this depends on the lattice vectors only.

As mentioned in the main text, we use finite samples instead of periodically repeated crystal structures to explicitly prove the local structure recognition capabilities of the method. 
Each system is then isotropically scaled by its average atomic bond length (i.e. distance between nearest neighboring atoms).
We also notice that for materials formed by hydrogen or helium the diffraction fingerprint contrast is low due to the small $f_a^{\lambda}$ (Eq. \ref{eq:psi_q}) of these elements; H and He are indeed notoriously difficult to detect with x-ray diffraction methods because of their small number of electrons ($Z=1$ and $Z=2$, respectively)\cite{de2007structure}.
However, our main goal here is to introduce a transferable descriptor for crystal structure representation, and not to compare with experimental data. Thus, we are free to choose a different value for the atomic number in order to augment the contrast in the diffraction fingerprint. In particular, we increase the atomic number of the elements by two when calculating the diffraction fingerprint, i.e. H is mapped to Li, He to Be, etc.
Moreover, given that the task is to distinguish crystals classes with an image for each system, one needs to choose a wavelength which is much smaller than the spacing between atoms, such that many beams are diffracted simultaneously (because the corresponding Ewald sphere radius is much larger than the lattice spacing)\cite{de2007structure}.
Therefore, we use a wavelength of $\lambda = 5.0 \cdot 10^{-12}$m for the incident plane wave (Eq.\ref{eq:psi_q}), a wavelength typically used in electron diffraction experiments. 
Indeed, the two-dimensional diffraction fingerprint bears resemblance to experimental scattering techniques such as single-crystal or selected-area electron diffraction; from this perspective, the angle of rotation could be chosen based on specific crystal orientations\cite{Bunge1982,Britton2016}.

For the (computational) detector, we use a pixel width and height of $4.0 \cdot 10^{-4}$m, and produce a 64$\times$64 pixel image as diffraction fingerprint.
Since the direct beam does not carry any structural information, and gives raise to a very bright central diffraction spot which compromises the contrast of high-order peaks, we remove this central spot from the diffraction fingerprint setting to zero the intensity within a radius of five pixels from the image center. The two-dimensional diffraction patterns are calculated using the open-source software Condor\cite{Hantke2016}.

\subsection*{Dataset}
Our pristine dataset consists of materials from the AFLOWLIB elemental solid database\cite{Curtarolo2012} belonging to centrosymmetric space groups which are represented with more than 50 configurations in the database. 
Specifically, we extract structures that have a consistent space group classification for different symmetry tolerances, as determined by the Python Materials Genomics (pymatgen)\cite{Ong2013} wrapper around the Spglib\cite{Spglib2017} library
 with {\tt symprec}$=\{10^{-3}$\AA, $10^{-6}$\AA, $10^{-9}$\AA $\}$ for all except rh and hex structures, 
for which {\tt symprec}$=\{10^{-3}$\AA, $10^{-6}$\AA $\}$ is employed since some symmetries are missed for {\tt symprec}=$10^{-9}$\AA.
This gives us crystal structures belonging to the following space groups: 139 (bct), 141 (bct), 166 (rh), 194 (hex), 221 (sc), 225 (fcc), 227 (diam), and 229 (fcc).
From this, we apply the defective transformations described in the main text (random displacements, vacancies, and chemical substitutions) to the pristine structures; the resulting dataset is used as test set. 
For this defective dataset we use labels from the pristine structures because the materials' class will unlikely be changed by the transformations above.
To quantify this, let us consider the transformation of bcc into sc crystals for the case of random vacancies as illustrative example. 
As stated in the main text, a sc structure can be obtained removing all atoms laying at the center of the bcc unit cell (see Fig.\ref{fig:descriptor}b). Therefore, for a structure comprising $N$ atoms, one needs to remove exactly the $N/2$ atoms which are at the center of the cubic unit cell (note that each corner atom is shared equally between eight adjacent cubes and therefore counts as one atom). For $N/2$ randomly generated vacancies, the probability of removing all and only these central atoms is $P_N = 2\left[\binom{N}{N/2}\right]^{-1}$ which - for the structure sizes considered in this work - leads to negligible probabilities ($P_{64} \approx 10^{-18}$, $P_{128} \approx 10^{-38}$). 
The same holds for chemical substitutions: even if in principle they could change the space group (e.g. diamond to zincblende structure), the probability of this to happen is comparable with the example above, and therefore negligible.
Finally, in the case of displacements, atoms are randomly moved about their original positions, and - due to this randomness - it is not possible to obtain any long-range re-organization of the crystal, necessary to change the materials' class; moreover, for large displacements the system becomes amorphous (without long-range order). 

\subsection*{Neural network architecture and training procedure}

The architecture of the convolutional neural network used in this work is detailed in Table \ref{tab:convnet_architecture}. Training was performed using Adam optimization\cite{Kingma2014} with batches of 32 images for 5 epochs with learning rate $10^{-3}$, and cross-entropy as cost function.
The convolutional  neural network was implemented with TensorFlow\cite{tensorflow2015-whitepaper} and Keras\cite{chollet2015}.

\begin{table}[ht]
\centering
\begin{tabular}{ll}
\hline
Layer type & Specifications \\
\hline
\hline
Convolutional Layer & (Kernel: 7x7; 32 filters)\\
Convolutional Layer & (Kernel: 7x7; 32 filters)\\
Max Pooling Layer & (Pool size: 2x2, stride: 2x2) \\
Convolutional Layer & (Kernel: 7x7; 16 filters)\\
Convolutional Layer & (Kernel: 7x7; 16 filters)\\
Max Pooling Layer & (Pool size: 2x2, stride: 2x2) \\
Convolutional Layer & (Kernel: 7x7; 8 filters)\\
Convolutional Layer & (Kernel: 7x7; 8 filters)\\
Fully connected Layer + Dropout & (Size: 128; dropout: 25\%) \\
Batch Normalization & (Size: 128) \\
Softmax & (Size: 7) \\
\hline
\end{tabular}
\caption{\label{tab:convnet_architecture}Architecture of the convolutional neural network used in this work.}
\end{table}

\subsection*{Data availability}
Calculation data can be downloaded from the Novel Materials Discovery (NOMAD) Repository and Archive (https://www.nomad-coe.eu/); the uniform resource locators (URLs) are provided in the Supplementary Note 3. Additional data including spatial coordinates and diffraction fingerprint for each structure of the pristine dataset is available at the Harvard Dataverse: \url{https://doi.org/10.7910/DVN/ZDKBRF}.
An online tutorial\cite{ziletti2018} to reproduce the main results presented in this work can be found in the NOMAD Analytics-Toolkit.


\section*{Acknowledgements}
A.Z., L.M.G., and M.S. acknowledge funding from the European Union's Horizon 2020 research and innovation programme, Grant Agreement No. 676580 through the Novel Materials Discovery (NOMAD) Laboratory, a European Center of Excellence (https://www.nomad-coe.eu). D.K would like to thank Dr. Vlado Menkovski for helpful discussions regarding visualization.
 
\section*{Author contributions}
A.Z., M.S., and L.M.G. conceived the project.  A.Z. performed the calculations. A.Z. and D.K. carried out the classification model visualization. A.Z, M.S., and L.M.G. wrote the manuscript.  All authors reviewed and commented on the manuscript.

\section*{Competing interests}
The authors declare no competing interests.

\section*{Additional information}
The authors declare that they have no competing financial interests.

\newpage

{\large Supplementary Information for:} \newline
\begin{center}
\title{\textbf{\LARGE Insightful classification of crystal structures using deep learning}}
\end{center}

\beginsupplement

\author{Angelo Ziletti}\textsuperscript{1,*}
\author{Devinder Kumar}\textsuperscript{2}
\author{Matthias Scheffler}\textsuperscript{1}
\author{Luca M. Ghiringhelli}\textsuperscript{1} 
\vspace{3em} 
\newline
\textsuperscript{1} Theory Department, Fritz-Haber-Institut der Max-Planck-Gesellschaft, Faradayweg 4-6, D-14195 Berlin, Germany \newline
\textsuperscript{1} University of Waterloo, 200 University Avenue West, N2L 3G1 Waterloo-Ontario, Canada
\newline
\textsuperscript{*} ziletti@fhi-berlin.mpg.de
\newline

\newpage
\section{Supplementary Note 1}
Here, we show a complete comparison between a state-of-the-art method (Spglib) and the proposed approach in classifying defective structures when random displacements are applied (Supplementary Table \ref{table:accuracy-comparison-disp}), or vacancies are created (Supplementary Table \ref{table:accuracy-comparison-vac}), as described in the main text.
The thresholds indicated as ``tight", ``medium", ``loose" in the tables below (and in Table \ref{table:accuracy-comparison} in the main text) are obtained with the following choices of Spglib parameters: 
``tight"= $\lbrace${\tt symprec}$=1 \times 10^{-3}$\AA, {\tt angle\_tolerance}$=1^{\circ}$ $\rbrace$; 
``medium"= $\lbrace${\tt symprec}$=1 \times 10^{-2}$\AA, {\tt angle\_tolerance}$=1^{\circ}$ $\rbrace$;
``loose"=$\lbrace${\tt symprec}$=1 \times 10^{-1}$\AA, {\tt angle\_tolerance}$=5^{\circ}$ $\rbrace$.
Finally, concerning defective structures obtained by chemical substitution, we find that our method is always able to identify the correct crystal class.

\begin{table*}[h]
\centering
\ra{1.5}
\begin{tabular}{@{}c p{0.9cm}p{0.9cm}p{0.9cm}p{0.9cm}p{0.9cm}p{0.9cm}p{0.9cm}p{0.9cm}p{0.9cm}p{0.9cm}p{0.9cm} c@{}} \hline \hline
& \multicolumn{11}{c}{Random Displacements ($\sigma$)} \\ 
\cline{2-12}
&  $0.001$\AA & $0.002$\AA &  $0.003$\AA & $0.004$\AA & $0.005$\AA 
&  $0.01$\AA &  $0.02$\AA &  $0.04$\AA & $0.06$\AA & $0.08$\AA & $0.10$\AA 
\\ \hline
Spglib (tight) & 0.00 & 0.00 & 0.00 & 0.00 & 0.00 & 0.00 
& 0.00 & 0.00 & 0.00 & 0.00 & 0.00 
\\
Spglib (medium) & 73.70 & 0.00 & 0.00 & 0.00 & 0.00 & 0.00 
& 0.00 & 0.00 & 0.00 & 0.00 & 0.00  
\\
Spglib (loose) & 99.99 & 99.99 & 99.99 & 99.99 & 99.99 & 75.22 
& 0.00 & 0.00 & 0.00 & 0.00 & 0.00  
\\
\textbf{This work} & 100.00 & 100.0 & 100.00 & 100.00 & 100.00 & 100.00 
& 100.00 & 100.00 & 100.00 & 99.97 & 99.91 
\\
\hline \hline
\end{tabular}
\caption{Accuracy in identifying the correct (most similar) crystal class in the presence of defects. The defective structures are calculated randomly displacing atoms according to Gaussian distribution with standard deviation $\sigma$. The accuracy values are in percentage.}
\label{table:accuracy-comparison-disp}
\end{table*}

\begin{table*}[h]
\centering
\ra{1.5}
\begin{tabular}{@{}c p{0.82cm}p{0.82cm}p{0.82cm}p{0.82cm}p{0.82cm}p{0.82cm}p{0.82cm}p{0.82cm}p{0.82cm}p{0.82cm}p{0.82cm}p{0.82cm}p{0.82cm}p{0.82cm} c@{}}\hline \hline
& \multicolumn{12}{c}{Vacancies ($\eta$)} \\ 
\cline{2-13}
&  1 \% & 2 \% & 5 \% & 10 \% & 15 \% & 20 \% & 25 \%
&  30 \% & 40 \% & 50 \% & 60 \% & 70 \% 
\\ \hline
Spglib (tight) & 0.02 & 0.00 & 0.00 & 0.00 & 0.00 
& 0.00 & 0.00 & 0.00 & 0.00 & 0.00 & 0.00 & 0.00 
\\
Spglib (medium) & 0.02 & 0.00 & 0.00 & 0.00 & 0.00 
& 0.00 & 0.00 & 0.00 & 0.00 & 0.00 & 0.00 & 0.00 
\\
Spglib (loose) & 0.01 & 0.00 & 0.00 & 0.00 & 0.00 
& 0.00 & 0.00 & 0.00 & 0.00 & 0.00 & 0.00 & 0.00 
\\
\textbf{This work} & 100.00 & 100.00 & 100.00 & 100.00 & 100.00 
& 100.00 & 100.00 & 100.00 & 100.00 & 99.56 & 97.18 & 89.05
\\
\hline \hline
\end{tabular}
\caption{Accuracy in identifying the correct (most similar) crystal class in the presence of defects. The defective structures are calculated randomly removing $\eta$\% of the atoms, thus creating vacancies. The accuracy values are in percentage.}
\label{table:accuracy-comparison-vac}
\end{table*}

\section{Supplementary Note 2}

\begin{figure}[ht]
\centering
\includegraphics[width=18cm]{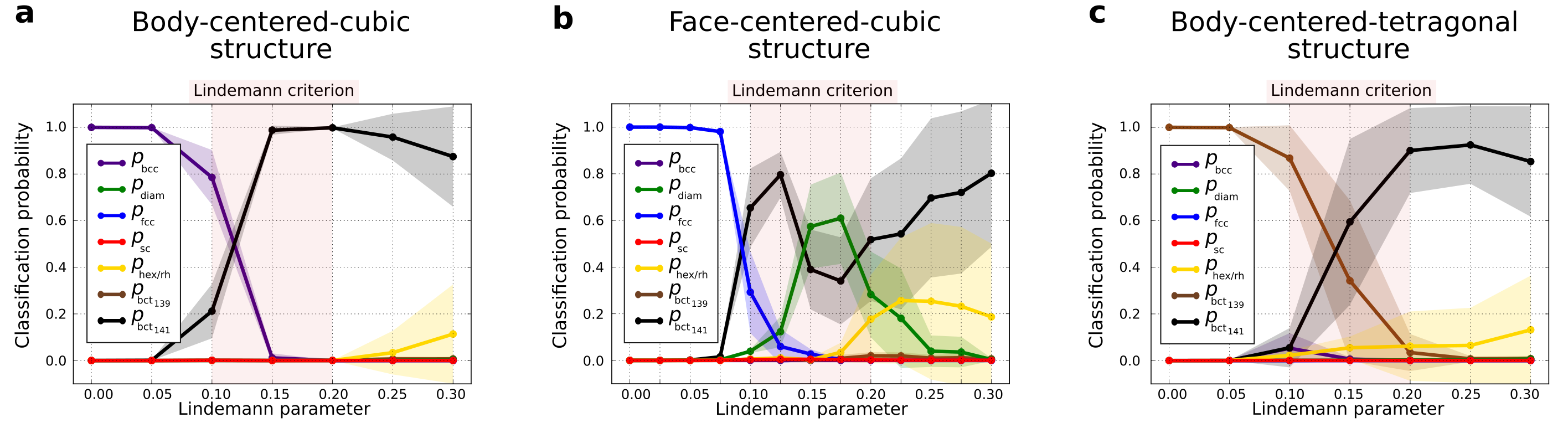}
\caption{From ordered to amorphous structures: the case of (a) fcc, (b) bcc, and (c) bct$_{139}$ crystal structures. The distributions of classification probability of all crystal classes are shown in each plot. The drop in the classification probability distribution for the fcc (a), bcc (b), and bct$_{139}$ (c) occurs around to the range of commonly accepted Lindemann parameters (shaded pink area between 0.1 and 0.2) for a solid to liquid transition (see text for more details).}
\label{fig:lindemann_transition}
\end{figure}

Since in the main text we have shown that the diffraction descriptor is well-behaved with respect to defects (Fig. \ref{fig:descriptor}), and that the neural network can extrapolate correctly to intermediate structures (see Fig. \ref{fig:defects}a), let us now address the question if our method can distinguish ordered structures with defects and amorphous structures.
There is an empirical (universal) criterion - the so-called Lindemann melting rule\cite{S_Lindemann1910} - to predict the solid-liquid transition: it states that this occurs when the ratio between the root-mean-square fluctuations about lattice positions and the nearest neighbor distance (termed Lindemann parameter) exceeds a certain threshold.
Here, we investigate whether structures, to which random displacements to the atomic positions are applied, are recognized as amorphous structures when the resulting Lindemann parameter exceeds the threshold. In practice, we generate structures with Lindemann parameter ranging from 0 (pristine) to 0.3 (amorphous) from a subset of fcc, bcc, and bct$_{139}$ structures belonging to the pristine dataset. 
In Supplementary Fig. \ref{fig:lindemann_transition} we plot the classification probability of all classes versus Lindemann parameter.
Let us here discuss in detail the case of fcc shown in Supplementary Fig. \ref{fig:lindemann_transition}a. Up to a Lindemann parameter of 0.075, the model classifies these disordered structures as fcc with probability close to 1, and zero standard deviation. In the range 0.1-0.2, however, the fcc probability classification drops significantly, just as the standard deviation increases: the neural network is now less confident on its predictions.  
For Lindemann parameters greater than 0.2 the structures are amorphous, and therefore the neural network classification is no longer meaningful: this is reflected in the very large standard deviation of the classification probability distributions. It is interesting to note that the drop of confidence in the neural network predictions falls in the range of commonly accepted critical values for the Lindemann parameter (0.1-0.2) \cite{S_young1991phase}. 
This is interesting, especially given that our model was built using exclusively pristine structures, and has thus no information regarding disordered structures. 
The situation for bcc and bct$_{139}$ classes (shown in Supplementary Fig. \ref{fig:lindemann_transition}b and \ref{fig:lindemann_transition}c, respectively) follow a quantitative different but qualitatively comparable trend.

\section{Supplementary Note 3}
The raw data used to generate the pristine dataset (corresponding to the AFLOWLIB elemental solid database)  can be downloaded from NOMAD Repository with the following links:
\begin{enumerate}[noitemsep]
\item \url{http://data.nomad-coe.eu/raw-data/data/Rnh/Rnh_4DFTJQgTSOib4e4d-5GByiTVB.zip}
\item \url{http://data.nomad-coe.eu/raw-data/data/R10/R10ncY1AZG6X9y-Nj8F0_DiN8NeLD.zip}
\item \url{http://data.nomad-coe.eu/raw-data/data/RsL/RsLoZhSAdK0BopfI2T4B5pLfMyjVN.zip}
\item \url{http://data.nomad-coe.eu/raw-data/data/RMG/RMGpPc3B_HiR0D-oLE4ND66HmYdH-.zip}
\item \url{http://data.nomad-coe.eu/raw-data/data/Re2/Re2mnhOAs6ZNqvTY1p-W2RavinjOM.zip}
\item \url{http://data.nomad-coe.eu/raw-data/data/R9u/R9usAWjw2xq9F8zW-66jyCyeDLlDa.zip}
\item \url{http://data.nomad-coe.eu/raw-data/data/Rkx/RkxmUCgPxt-9xDdIpr5xqPQK8PC9H.zip}
\item \url{http://data.nomad-coe.eu/raw-data/data/Rdz/RdzeezGR0W5wGEpGYEqOq7AygYS9J.zip} 
\item \url{http://data.nomad-coe.eu/raw-data/data/Rc_/Rc_XxYadb0ZlfBVLqCNo-EtVocxv8.zip} 
\item \url{http://data.nomad-coe.eu/raw-data/data/RKX/RKXqE9xPCiLlufNK0n4pbtzdbID5H.zip} 
\item \url{http://data.nomad-coe.eu/raw-data/data/Ryv/RyvdvBLf1QdM5QJ_8DVve7CknkdK5.zip} 
\item \url{http://data.nomad-coe.eu/raw-data/data/Rut/Rut3qcReY6SJO6fIJ5jangTSlMjaQ.zip}   
\item \url{http://data.nomad-coe.eu/raw-data/data/Reg/Reg0D-KojGnrw51EYl2Q2rCYOIfJM.zip}  
\item \url{http://data.nomad-coe.eu/raw-data/data/RSk/RSkoltrNkpZwp1xpi_Oj1jO4IndC5.zip}  
\item \url{http://data.nomad-coe.eu/raw-data/data/RA-/RA-tqhSlH5idfPW_3UxE80I7BBL6s.zip}  
\item \url{http://data.nomad-coe.eu/raw-data/data/Ray/RayT1o-XjyZaWdlVS_Fk8nssdO1w9.zip}  
\item \url{http://data.nomad-coe.eu/raw-data/data/Rlb/RLbbgx7klbZ7ZdO5O_YABQGjBOZ9g.zip}  
\item \url{http://data.nomad-coe.eu/raw-data/data/RcC/RcC8TDWGWCtQLhWeB2a1N8y9Q7y4r.zip}  
\item \url{http://data.nomad-coe.eu/raw-data/data/Ra8/Ra8nAuJOgxGwSytw1scU5BTeB3ozo.zip}
\end{enumerate}

Additional data including atomic structures and two-dimensional diffraction fingerprints of the pristine dataset are available from the Harvard Dataverse: \url{https://doi.org/10.7910/DVN/ZDKBRF}.


\end{document}